\documentclass[conference,letterpaper]{IEEEtran}

\usepackage[utf8]{inputenc} 
\usepackage[T1]{fontenc}

\usepackage{dsfont}
\usepackage{units}

\usepackage{url}              
\usepackage{cite}             

\usepackage[cmex10]{amsmath}  
\usepackage{mleftright}       
\mleftright                   

\usepackage{graphicx}         
\usepackage{booktabs}         

\usepackage{amsmath,cite,amsfonts,amssymb,psfrag,paralist}
\usepackage{epstopdf}
\usepackage{rotating}
\usepackage{dsfont}
\usepackage{color}
\usepackage{tikz}
\usetikzlibrary{plotmarks}
\usepackage{pgfplots}
\usetikzlibrary{calc}
\usetikzlibrary{shapes,arrows}
\usetikzlibrary{decorations.markings}
\usetikzlibrary{positioning}
\pgfplotsset{compat=1.10}
\usetikzlibrary{calc}
\usetikzlibrary{shapes,arrows}
\usetikzlibrary{decorations.markings}

\usepackage{balance}

\usepackage[utf8]{inputenc}

\def\BibTeX{{\rm B\kern-.05em{\sc i\kern-.025em b}\kern-.08em T\kern-.1667em\lower.7ex\hbox{E}\kern-.125emX}}

\DeclareFontFamily{U}{mathx}{\hyphenchar\font45}
\DeclareFontShape{U}{mathx}{m}{n}{<-> mathx10}{}
\DeclareSymbolFont{mathx}{U}{mathx}{m}{n}

\newcommand{\Enc}{\mathsf{Enc}}
\newcommand{\Dec}{\mathsf{Dec}}
\newcommand{\Est}{\mathsf{Est}}

\usepackage{scalerel}
\usepackage{stackengine}
\stackMath
\def\hatgap{2pt}
\def\subdown{-2pt}
\newcommand\reallywidehat[2][]{%
	\renewcommand\stackalignment{l}%
	\stackon[\hatgap]{#2}{%
		\stretchto{%
			\scalerel*[\widthof{$#2$}]{\kern-.6pt\bigwedge\kern-.6pt}%
			{\rule[-5\textheight]{0.1ex}{\textheight}}
		}{0.5ex}
		_{\smash{\belowbaseline[\subdown]{\scriptstyle#1}}}%
}}

\newtheorem{theorem}{Theorem}

\newtheorem{definition}{Definition}
\newtheorem{proposition}{Proposition}

\newtheorem{corollary}{Corollary}

\newcommand{\R}{\mathds{R}}
\newcommand{\E}{\mathds{E}}
\newcommand{\rsym}{\ensuremath{s}}
\newcommand{\Rsym}{\ensuremath{S}}

\newcommand{\rvRx}{\ensuremath{\Rsym_1}}

\newcommand{\rvRy}{\ensuremath{\Rsym_2}}

\newcommand{\srx}{\ensuremath{\sigma_{\rvRx}}}
\newcommand{\sry}{\ensuremath{\sigma_{\rvRy}}}
\newcommand{\srxq}{\ensuremath{\srx^2}}
\newcommand{\sryq}{\ensuremath{\sry^2}}

\newcommand{\fcor}{\ensuremath{\rho^2}}

\newcommand{\bcor}{\ensuremath{\tilde{c}}}
\newcommand{\sxtd}{\ensuremath{\tilde{\sigma}_1^2}}

\newcommand{\Ralpha}{\ensuremath{R_{\alpha}}}
\newcommand{\RalphaUB}{\ensuremath{R_{\alpha,\mathrm{ub}}}}
\newcommand{\Rbeta}{\ensuremath{R_{\beta}}}

\begin{document}

\title{Secure Integrated Sensing and Communication Under Correlated Rayleigh Fading} 

	\IEEEoverridecommandlockouts
	
	\author{%
		\IEEEauthorblockN{Martin          
            Mittelbach\textsuperscript{1},
            Rafael F. Schaefer\textsuperscript{1},
			Matthieu Bloch\textsuperscript{2},
			Aylin Yener\textsuperscript{3},
            and Onur G\"unl\"u\textsuperscript{4}
		}
		\IEEEauthorblockA{\textsuperscript{1}%
			Chair of Information Theory and Machine Learning, TU Dresden, 
			\{martin.mittelbach, rafael.schaefer\}@tu-dresden.de
		}
		\IEEEauthorblockA{\textsuperscript{2}%
			School of Electrical and Computer Engineering, Georgia Institute of Technology, matthieu.bloch@ece.gatech.edu
		}
		\IEEEauthorblockA{\textsuperscript{3}%
			Department of Electrical and Computer Engineering, The Ohio State University, yener@ece.osu.edu
		}
  	\IEEEauthorblockA{\textsuperscript{4}%
			Information Theory and Security Laboratory, Link{\"o}ping University, 
			onur.gunlu@liu.se
		}
	}

\maketitle

\begin{abstract}
      	We consider a secure integrated sensing and communication (ISAC) scenario, in which a signal is transmitted through a state-dependent wiretap channel with one legitimate receiver with which the transmitter communicates and one honest-but-curious target that the transmitter wants to sense. The secure ISAC channel is modeled as two state-dependent fast-fading channels with correlated Rayleigh fading coefficients and independent additive Gaussian noise components. Delayed channel outputs are fed back to the transmitter to improve the communication performance and to estimate the channel state sequence. We establish and illustrate an achievable secrecy-distortion region for degraded secure ISAC channels under correlated Rayleigh fading. We also evaluate the inner bound for a large set of parameters to derive practical design insights for secure ISAC methods. The presented results include in particular parameter ranges for which the secrecy capacity of a classical wiretap channel setup is surpassed and for which the channel capacity is approached.
\end{abstract}

\section{Introduction}\label{sec:introduction}
Integrating the digital and physical world, envisioned for future communication systems, requires a network to react to changes in real-time through sensing and communication \cite{NokiaGuysJCASTutorial}. An insightful example is a millimeter wave (mmWave) integrated sensing and communication (ISAC) system that aims to sense a target by estimating relevant channel parameters to fine-tune the communication scheme \cite{JCASwithSecurityTutorial,VDEJCASPositionPaper}.

There have recently been multiple information-theoretic studies of ISAC that extend previous results, such as  \cite{Zhang_2011,AcademicsJCASTutorial,MassiveMIMOforJCAS}. Focusing on vehicular radar applications for mmWave systems, an information-theoretic model is proposed in \cite{MariMichelleGJournalBroadcast} for ISAC. In this model, encoded messages are sent over a state dependent channel with generalized feedback such that the state is  only known at the receiver and the feedback is used to improve communication and to estimate the channel state. The rate-distortion region is characterized for independent and identically distributed (i.i.d.) channel states and memoryless ISAC channels with strictly causal channel output feedback. Subsequent works include extensions to multiple access channels~\cite{MariMACISIT}, broadcast channels~\cite{MariMichelleGJournalBroadcast}, two-way channels \cite{MichelleCollaborative}, and transmitter actions \cite{TrumanISIT2024}.

Since a single modality is used to both communicate with a legitimate receiver and detect a target, the sensing signal may carry sensitive information about the message communicated, which may then is leaked to a curious target. Since the signal power at the sensed target impacts both the secrecy and sensing performance, there exists  a tradeoff between the two \cite{oursecureISACJSAIT,JCASwithSecurityTutorial,OurJCandSConferenceISAC, MichelleSecureISACwithSholomo}. This tradeoff is characterized in \cite{oursecureISACJSAIT} for degraded and reversely-degraded ISAC channels, when the transmitter aims to reliably communicate with the legitimate receiver by using the ISAC channel, estimate the channel state by using the channel output feedback, and keep the message hidden from the target that acts as an eavesdropper. The results in \cite{oursecureISACJSAIT} show that it is possible to surpass the secrecy capacity by using the channel output feedback for secure ISAC applications, which is in line with the insights from wiretap channel with feedback results, such as  in \cite{AhlswedeCaiWTCwithFeedback, AsafCohenWTCwithFeedback,OurJSAITTutorial, HanVinckWTCwithFeedback,he-yener-fbsecrecy, GermanWTCwithGeneralizedFeedback}.

In this work, we establish an achievable rate region for stochastically-degraded secure ISAC channels under bivariate Rayleigh fading by using a Gaussian channel input. Since closed form expressions for this rate region remain elusive, we derive integral expressions from the involved differential entropies, which are amenable to simplified and stable numerical evaluations. Based on the evaluation results, insights are presented including, in particular, parameter ranges for which secure-ISAC rates greater than the secrecy capacity can be achieved and for which the channel capacity is approached. Moreover, we provide accurate approximations, which allow easy-to-compute numerical evaluations.

\section{System Model and Problem Definition}\label{sec:problem_setting}

\begin{figure*}
	\centering
	\resizebox{0.99\linewidth}{!}{
		\begin{tikzpicture}
		\node (message) at (0,0) {$M$};
		\node (enc) [right of = message, node distance = 2.5cm] [draw,rounded corners = 5pt, minimum width=1cm,minimum height=0.6cm, align=left] {$X_i=\Enc_i(M,Z^{i-1})$};
		\node (xi) [right of = enc, node distance = 3cm] {$X_i$};
		\node (est) [below of = xi, node distance = 2.5cm] [draw,rounded corners = 5pt, minimum width=1cm,minimum height=0.6cm, align=left] {$\widehat{\Rsym^n_j} = \Est_j(X^n,Z^n)$};
		\draw[decoration={markings,mark=at position 1 with {\arrow[scale=1.5]{latex}}}, postaction={decorate}, thick, shorten >=1.4pt]  (xi.south) -- (est.north);
		\draw[decoration={markings,mark=at position 1 with {\arrow[scale=1.5]{latex}}}, postaction={decorate}, thick, shorten >=1.4pt]  (enc.east) -- (xi.west);
		\draw[decoration={markings,mark=at position 1 with {\arrow[scale=1.5]{latex}}},
		postaction={decorate}, thick, shorten >=1.4pt] (message.east) -- (enc.west) ;
		\node (s1s2) [right of = xi, node distance = 2.5cm] [draw,rounded corners = 5pt, minimum width=1cm,minimum height=0.6cm, align=left]{$f_{\Rsym_1\Rsym_2}$};
		\node (multiple1) at (8,2) {$\bigotimes$};
		\node (multiple2) at (8,-2) {$\bigotimes$};
		\draw[decoration={markings,mark=at position 1 with {\arrow[scale=1.5]{latex}}}, postaction={decorate}, thick, shorten >=1.4pt]  (s1s2.north) -- ($(s1s2.north)+(0.0,0.3)$) --  ($(multiple1.south) + (0, -1.13)$) -- ($(multiple1.south)+(0,0.12)$) node [near end, right, yshift=-0.25cm, xshift=-0.1cm] {$\Rsym_{1,i}$};
		\draw[decoration={markings,mark=at position 1 with {\arrow[scale=1.5]{latex}}},postaction={decorate}, thick, shorten >=1.4pt]  (s1s2.south) -- ($(s1s2.south)+(0.0,-0.3)$) --  ($(multiple2.north) + (0, 1.1)$) -- ($(multiple2.north)+(0,-0.1)$) node [near end, right, yshift=0.25cm, xshift=-0.1cm] {$\Rsym_{2,i}$};
		\draw[decoration={markings,mark=at position 1 with {\arrow[scale=1.5]{latex}}}, postaction={decorate}, thick, shorten >=1.4pt]  (xi.east) -- ($(multiple1.south)+(-0.1,0.12)$);
		\draw[decoration={markings,mark=at position 1 with {\arrow[scale=1.5]{latex}}}, postaction={decorate}, thick, shorten >=1.4pt]  (xi.east) -- ($(multiple2.north)+(-0.15,-0.1)$);
		\node (sum1) [right of = multiple1, node distance = 3cm]  {$\bigoplus$};
		\node (sum2) [right of = multiple2, node distance = 3cm]{$\bigoplus$};
		\node (awgn1) [above of = sum1, node distance = 1cm]  {$N_{1,i}$};
		\node (awgn2) [below of = sum2, node distance = 1cm]{$N_{2,i}$};
		\draw[decoration={markings,mark=at position 1 with {\arrow[scale=1.5]{latex}}},postaction={decorate}, thick, shorten >=1.4pt]  ($(awgn1.south)+(0,0.1)$) -- ($(sum1.north)+(0,-0.1)$);
		\draw[decoration={markings,mark=at position 1 with {\arrow[scale=1.5]{latex}}},postaction={decorate}, thick, shorten >=1.4pt]  ($(awgn2.north)+(0,-0.1)$) -- ($(sum2.south)+(0,0.1)$);
		\draw[decoration={markings,mark=at position 1 with {\arrow[scale=1.5]{latex}}}, postaction={decorate}, thick, shorten >=1.4pt]  ($(multiple1.east)-(0.15,0)$) -- ($(sum1.west)+(0.15,0)$);
		\draw[decoration={markings,mark=at position 1 with {\arrow[scale=1.5]{latex}}}, postaction={decorate}, thick, shorten >=1.4pt]  ($(multiple2.east)-(0.15,0)$) -- ($(sum2.west)+(0.15,0)$);
		\node (dec) [right of = sum1, node distance = 4cm] [draw,rounded corners = 5pt, minimum width=1cm,minimum height=0.6cm, align=left] {$\widehat{M}=\Dec(Y_1^n,\Rsym_1^n)$};
		\node (mhat) [right of = dec, node distance = 2.5cm] {$\widehat{M}$};
		\draw[decoration={markings,mark=at position 1 with {\arrow[scale=1.5]{latex}}}, postaction={decorate}, thick, shorten >=1.4pt]  (dec.east) -- (mhat.west);
		\node (eve) [right of = sum2, node distance = 4cm] [draw,rounded corners = 5pt, minimum width=1cm,minimum height=0.6cm, align=left] {EVE};
		\draw[decoration={markings,mark=at position 1 with {\arrow[scale=1.5]{latex}}}, postaction={decorate}, thick, shorten >=1.4pt]  ($(sum1.east)-(0.15,0)$) -- ($(dec.west)+(0.0,0)$)  node [midway, above] {$Y_{1,i}$};
		\draw[decoration={markings,mark=at position 1 with {\arrow[scale=1.5]{latex}}}, postaction={decorate}, thick, shorten >=1.4pt]  ($(sum2.east)-(0.15,0)$) -- ($(eve.west)+(0.0,0)$) node [midway, above] {$Y_{2,i}$};
		\draw[decoration={markings,mark=at position 1 with {\arrow[scale=1.5]{latex}}}, postaction={decorate}, thick, shorten >=1.4pt]  (s1s2.north) -- ($(s1s2.north)+(0.0,0.3)$) --  ($(dec.south) + (0, -1.02)$) -- ($(dec.south)+(0,0.00)$) node [near end, right, yshift=-0.25cm, xshift=-0.1cm] {$\Rsym_{1,i}$};
		\draw[decoration={markings,mark=at position 1 with {\arrow[scale=1.5]{latex}}}, postaction={decorate}, thick, shorten >=1.4pt]  (s1s2.south) -- ($(s1s2.south)+(0,-0.3)$) --  ($(eve.north) + (0, 1.1)$) -- ($(eve.north)+(0,0)$) node [near end, right, yshift=0.25cm, xshift=-0.1cm] {$\Rsym_{2,i}$};
		\node (delay) [right of = s1s2, node distance = 10cm] [draw,rounded corners = 5pt, minimum width=1cm,minimum height=0.6cm, align=left] {Delay};
		\draw[decoration={markings,mark=at position 1 with {\arrow[scale=1.5]{latex}}}, postaction={decorate}, dashed, shorten >=1.4pt]  ($(sum1.east)+(0.4,0)$) -- ($(sum1.east)+(0.4,-1.9)$) --  ($(delay.west) + (0, 0.1)$) node [near end, above, yshift=0.0000025cm, xshift=0.4cm] {$Y_{1,i}$};
		\draw[decoration={markings,mark=at position 1 with {\arrow[scale=1.5]{latex}}}, postaction={decorate}, dashed, shorten >=1.4pt]  ($(sum2.east)+(0.4,0)$) -- ($(sum2.east)+(0.4,+1.9)$) --  ($(delay.west) + (0, -0.1)$) node [near end, below, yshift=0.0000025cm, xshift=0.4cm] {$Y_{2,i}$};
		\draw[decoration={markings,mark=at position 1 with {\arrow[scale=1.5]{latex}}}, postaction={decorate}, dashed, shorten >=1.4pt]  ($(delay.south)+(0,0)$) -- ($(delay.south)+(0,-3.3)$) --  ($(est.south) + (0, -0.8)$) --  ($(est.south) + (0, 0)$) node [midway, below, yshift=-0.5cm, xshift=5.9cm] {$Z_{i-1} = f(Y_{1,i-1},Y_{2,i-1})$} node [near end, right, yshift=-0.15cm, xshift=0.2cm] {$Z_{i-1}$};
		\draw[decoration={markings,mark=at position 1 with {\arrow[scale=1.5]{latex}}}, postaction={decorate}, dashed, shorten >=1.4pt]  ($(delay.south)+(0,0)$) -- ($(delay.south)+(0,-3.3)$) --  ($(enc.south) + (0, -3.36)$) --  ($(enc.south) + (0, 0)$) node [near end, right, yshift=0.15cm, xshift=0.002cm] {$Z_{i-1}$};
		\end{tikzpicture}
	}
      \vspace*{-0cm}
	  \caption{Secure ISAC model for $i~=~[1:n]$ and $j=1,2$, for which the message $M$ should be kept secret from the eavesdropper. We impose an average transmit power constraint on the channel input symbols $X_i$ and assume independent AGN components $N_{1,i}$ and $N_{2,i}$. We principally consider perfect channel output feedback with unit symbol time delay, i.e., $Z_{i-1}=(Y_{1,i-1},Y_{2,i-1})$ such that the function $f(\cdot,\cdot)$ is the identity function.}\label{fig:SecureFadingISACModel}
	\vspace*{-0.1cm}
\end{figure*}
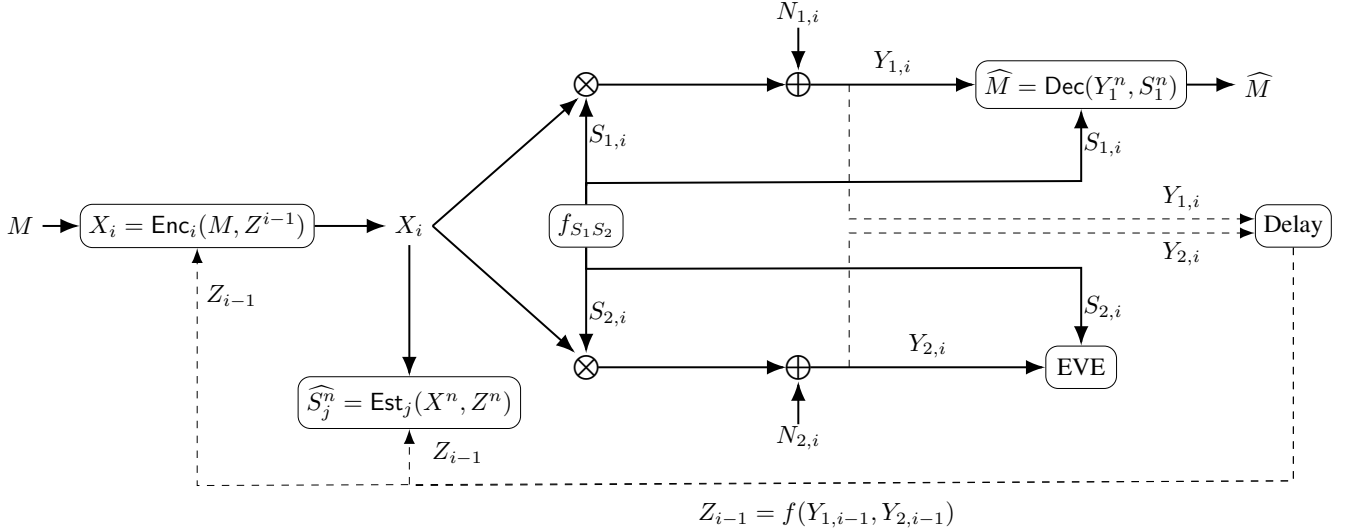

We consider the secure ISAC model depicted in Fig.\,\ref{fig:SecureFadingISACModel}, which includes one transmitter, one legitimate receiver, one state estimator, and an eavesdropper (Eve). The transmitter wants to transmit a uniformly distributed message $M$ from the finite message set $\mathcal{M}$ through a fast fading additive Gaussian noise (AGN) secure ISAC channel, in which i.i.d. fading channel coefficients $(\Rsym_1^n,\Rsym_2^n)$ are causally estimated by the receiver and eavesdropper, respectively. The fading coefficients $(\Rsym_{1,i},\Rsym_{2,i})$ with non-negative real-valued alphabet $\mathcal{\Rsym}_{1}\times\mathcal{\Rsym}_{2}$ are  correlated according to a known joint probability density function (pdf) $f_{\Rsym_1,\Rsym_2}$, but their realizations are not known by the transmitter. For discussions about how to extend the results to include complex fading channel coefficients and complex noise components, see \cite[Section~V-A]{TseFadingBC}.

Given $M$, the transmitter generates the channel inputs $X^n$ by using encoding functions $\Enc_i(\cdot)$ such that $X_i=\Enc_i(M,Z^{i-1})$ for all $i=[1:n]$, where $Z^{i-1}$ is the delayed channel output feedback. We impose an average power constraint on the subsequent transmitted symbols, i.e., we have
\begin{align}
		\frac{1}{n}\sum\nolimits_{i=1}^n\mathbb{E}[X_i^2]\leq P
\end{align} 
for all messages $M$, where $\mathbb{E}[\cdot]$ denotes expectation. The channel output for the legitimate receiver at time $i$ is
\begin{align}
	Y_{1,i} = \Rsym_{1,i}X_i+N_{1,i}\label{eq:Y1idef}
\end{align}
where $N_{1,i}$ are i.i.d. Gaussian distributed with zero mean, variance $\sigma_{N_1}^2$, and  independent of $(\Rsym_{1,i},\Rsym_{2,i},X_{i})$. The legitimate receiver observes the sequences $(Y_1^n,\Rsym_1^n)$ and estimates the transmitted message as $\widehat{M}=\Dec(Y_1^n,\Rsym_1^n)$, where $\Dec(\cdot)$ is a decoding function. Similarly, the channel output for the eavesdropper at time $i$ is 
\begin{align}
	Y_{2,i} = \Rsym_{2,i}X_i+N_{2,i}\label{eq:Y2idef}
\end{align}
where $N_{2,i}$ are i.i.d. Gaussian distributed with zero mean, variance $\sigma_{N_2}^2$, and independent of $(\Rsym_{1,i},\Rsym_{2,i},X_{i},N_{1,i})$. The transmitted message $M$ should be kept secret from the eavesdropper that observes $(Y_2^n,\Rsym_2^n)$. Finally, the state estimator observes both the channel output feedback
\begin{align}
	Z_{i-1}=f(Y_{1,i-1},Y_{2,i-1})
\end{align}
and the codeword symbol $X_i$ to estimate the fading channel coefficients $(\Rsym_1^n,\Rsym_2^n)$ as $\widehat{\Rsym^n_j} = \Est_j(X^n,Z^n)$ for $j=1,2$, where $\Est_j(\cdot,\cdot)$ is an estimation function with range ${\mathcal{\Rsym}_j^n}$. 

For simplicity, we assume the deterministic processing function $f(\cdot,\cdot)$ is the identity function, so the channel output feedback is perfect, i.e., we have noiseless channel output feedback $Z_{i-1}=(Y_{1,i-1},Y_{2,i-1})$. This simplification allows us to obtain fundamental insights into the optimal coding schemes and helps tackle the noisy feedback scenario, which is generally challenging; see, e.g., achievability results for wiretap channels with generalized output feedback in \cite{GermanWTCwithGeneralizedFeedback}. Note that the achievability proofs for wiretap channels generally require a local randomness source at the encoder, which is true also for the results given below. The randomness can be provided, e.g., by using hardware-intrinsic security primitives \cite{OurPUFTutorial}. We next define the secrecy-distortion region for the secure correlated fast-fading ISAC problem.

\begin{definition}\label{def:systemmodel}
	\normalfont \text{A secrecy-distortion tuple $(R,\!D_1,\!D_2)$ is \emph{achie}-} \emph{vable} for the secure correlated fast-fading ISAC problem if, for any $\epsilon\!>\!0$, there exist $n\!\geq\!1$, one encoder-decoder pair, and two state estimators $\Est_j(X^n,Y_j^n)=\widehat{\Rsym_j^n}$, $j=1,2$ such that
	\begin{align}
		&\Pr[M \neq \widehat{M}] \leq \epsilon&&\!\!\!\!\! (\text{reliability})\label{eq:reliability_cons}\\
		& \frac{1}{n}\log |\mathcal{M}|\geq R -\epsilon&&\!\!\!\!\!(\text{message rate})\label{eq:rates_cons}\\ 
		&\frac{1}{n}I(M;Y^n_2,\Rsym_2^n) \leq \epsilon&&\!\!\!\!\!(\text{weak secrecy})\label{eq:secrecyleakage_cons}\\
		&\mathbb{E}\big[d_j(\Rsym_j^n,\widehat{\Rsym_j^n})\big] \!\leq\! D_j\!+\!\epsilon\;\;\;\;\;\;\text{for } j\!=\!1,2\;\;&&\!\!\!\!\!(\text{distortions})\label{eq:distortion_consts}
	\end{align}
	where $d_j(\rsym^n,\widehat{\rsym^n})=\frac{1}{n}\sum_{i=1}^nd_j(\rsym_i,\widehat{\rsym}_i)$ for $j\!=\!1,2$ are averaged per-letter distortion metrics.
	
	The secrecy-distortion region $\mathcal{R}_{\text{S-ISAC}}$ is the closure of the set of all achievable tuples for the secure correlated fast-fading ISAC problem under perfect channel output feedback. \hfill $\lozenge$
\end{definition}

Since the transmitted message is independent of the channel state, the secrecy condition in (\ref{eq:secrecyleakage_cons}) is equivalent to the inequality $I(M;Y^n_2|\Rsym_2^n) \leq n\epsilon$.  Furthermore, there are ISAC models, such as in \cite{ISACMACChinesegroup}, that consider a practical application, in which only a part of the channel parameters are relevant for the transmitter. By not imposing the estimation of the exact channel state at the transmitter via adapting (\ref{eq:distortion_consts}), one can extend our results for such practical settings.

\section{Correlated-Fading AGN ISAC Channel Secrecy-Distortion Regions}
\subsection{Secrecy-Distortion Region}

We first define physically- and stochastically-degraded ISAC channels; see also \cite{MariMichelleGJournalBroadcast,oursecureISACJSAIT}.

\begin{definition}\label{def:physicallydegraded}
	\normalfont An ISAC channel is \emph{physically-degraded} if $X$ and $(Y_2,\Rsym_2)$ are conditionally independent given  $(Y_1,\Rsym_1)$.  
 Moreover, an ISAC channel is \emph{stochastically-degraded} if the joint probability distribution of $(X,Y_1,\Rsym_1,Y_2,\Rsym_2)$ can be preserved by using a marginal probability distribution of $(X,Y_1,\Rsym_1)$ such that the corresponding ISAC channel is physically-degraded. \hfill $\lozenge$
\end{definition}

We next state the secrecy-distortion region given in \cite{oursecureISACJSAIT} for physically-degraded secure ISAC channels under strong secrecy, i.e., $I(M;Y_2^n,\Rsym_2^n)\leq \epsilon$, for discrete-alphabet random variables and state estimators of the form $\Est_j(X^n,Y_1^n,Y_2^n)=\widehat{\Rsym_j^n}$. We then evaluate the entropy terms in the given rate region to characterize the secrecy-distortion regions for the secure ISAC channels considered in this work. 

\begin{theorem}[{\hspace{1sp}\cite[Theorem 1]{oursecureISACJSAIT}}]\label{theo:ourprevphydegrRegion}
	For physically-degraded ISAC channels, the secrecy-distortion region is the union w.\,r.\,t.\ all probability distributions $P_{X}$ of the rate-distortion tuples $(R,D_1,D_2)$ satisfying 
	\begin{align}
		&D_j\geq \mathbb{E}[d_j(\Rsym_j,\widehat{\Rsym}_j))]\qquad\qquad\qquad  \text{for }j=1,2\label{eq:distortionconstraintfordegraded}\\
		& R\leq \min\Big\{\big(H(Y_1,\Rsym_1|Y_2,\Rsym_2) - H(\Rsym_1|Y_1,Y_2,\Rsym_2,X)\big),\nonumber\\
		&\qquad\qquad\qquad\qquad I(X;Y_1|\Rsym_1)\Big\} \\
		&\Est_j(x,y_1,y_2)\nonumber\\
		&\qquad=\mathop{\textnormal{argmin}}_{\tilde{\rsym}\in\widehat{\mathcal{\Rsym}}_j} \sum_{\rsym_j\in\mathcal{\Rsym}_j}P_{\Rsym_j|XY_1Y_2}(\rsym_j|x,y_1,y_2)\; d_j(\rsym_j,\tilde{\rsym}).\label{eq:prevphydegrestimator}
	\end{align}
\end{theorem}

The measures in the secrecy-distortion region in Theorem~\ref{theo:ourprevphydegrRegion} remain valid for correlated-fading channels with independent AGN components, as depicted in Fig.~\ref{fig:SecureFadingISACModel}, for the following reasons: 
\begin{inparaenum}[(i)]
    \item any achievability proof for strong secrecy also follows under weak secrecy, and by replacing $\delta_n$ in \cite[Eq.~(72)]{oursecureISACJSAIT} with $n\delta_n$, one can obtain the matching converse for weak secrecy;  
	\item the outer bound applies to arbitrary random variables and does not assume any degradedness; 
	\item there is a discretization procedure to generalize the achievability proof to well-behaved continuous-alphabet random variables, such as the considered fading and noise distributions \cite[Remark~3.8]{Elgamalbook};  
	\item one can show that changing the estimator form does not change the entropy terms in the rate region, although achieved distortion levels might change since the estimators given in (\ref{eq:prevphydegrestimator}) should be also adapted.
\end{inparaenum}
Moreover, the state estimators considered in Definition~\ref{def:systemmodel} make the measures in Theorem~\ref{theo:ourprevphydegrRegion} valid also for stochastically-degraded channels, which follows because the constraints (\ref{eq:reliability_cons})--(\ref{eq:distortion_consts}) in Definition~\ref{def:systemmodel} only depend on the marginal probability distributions of $(X,Y_1,\Rsym_1)$ and $(X,Y_2,\Rsym_2)$. This extension is important, as the practical secure ISAC model considered in this work is not physically-degraded. 

We next consider the secrecy-distortion region for stochastically-degraded ISAC channels for the secure ISAC model given in Fig.\,\ref{fig:SecureFadingISACModel} and focus on the secrecy rate. In what follows expectations with subscripts of random variables indicate that we first calculate the argument of the expectation for fixed realization of the subscript and afterwards calculate the expectation w.\,r.\,t.\ the distribution of the subscript. Using \cite[Proposition~1]{OurJCandSConferenceISAC}, we have following result.

\begin{corollary}
The secrecy-distortion region for the secure ISAC model depicted in Fig.~\ref{fig:SecureFadingISACModel} is the union w.\,r.\,t.\ all pdfs $f_X$ of the rate-distortion tuples $(R,D_1,D_2)$  satisfying
\begin{subequations}\label{Eq:Achievable-Rate}
\begin{align}
        &R \leq\min\Big\{\E_{\Rsym_1,\Rsym_2}\big[h(\Rsym_1X\!+\!N_1|\Rsym_2X\!+\!N_2)\big] \!-\! h(N_1)\label{Eq:Achievable-Rate-a}\\
        &\qquad\qquad\qquad +\,\E_{X}\big[h(\Rsym_1X+N_1|\Rsym_2)\big],\label{Eq:Achievable-Rate-b}\\
        &\qquad\qquad\quad \E_{\Rsym_1}\big[h(\Rsym_1X+N_1)\big] -h(N_1)\Big\}\label{Eq:Achievable-Rate-c}
\end{align}        
\end{subequations}
such that (\ref{eq:distortionconstraintfordegraded}) is satisfied for a given distortion metric by using estimators of the form $\Est_j(X^n,Y_1^n,Y_2^n)=\widehat{\Rsym_j^n}$. 
\end{corollary}

Denote the complementary cumulative distribution function of a real-valued random variable $U$ as $\bar{F}_U(u)=\Pr[U\geq u]$. We provide a sufficient but not necessary condition to generate a stochastically-degraded secure ISAC channel based on the stochastic ordering of the channel outputs $Y_1$ and $Y_2$.

\begin{proposition}\label{prop:degrordering}
		The secure ISAC channel in (\ref{eq:Y1idef}) and (\ref{eq:Y2idef}) is stochastically-degraded if $\Rsym_1^2/\sigma_{N_1}^2$ is \emph{stochastically larger} than $\Rsym_2^2/\sigma_{N_2}^2$, i.e., if we have, for all $\rsym\geq 0$,
 		\begin{align}\label{eq:condition-stochastic-degradedness}
			\bar{F}_{\Rsym_1^2}\bigg(\frac{\rsym}{\sigma_{N_1}^2}\bigg)\geq  \bar{F}_{\Rsym_2^2}\bigg(\frac{\rsym}{\sigma_{N_2}^2}\bigg).
		\end{align}
\end{proposition}

The proof of Proposition~\ref{prop:degrordering} follows from \cite[Lemma~3]{TseFadingBC} after appropriate  changes to account for the noise variances. For necessity discussions see \cite[Lemmas~1--4]{PinHsunGaussianWiretap}.

We next specify the correlated fast-fading distribution, for which we characterize an achievable secrecy-distortion region, when the secure ISAC channel is stochastically-degraded.

\subsection{Bivariate Rayleigh Fading}
Suppose the fading random variables $(\Rsym_1,\Rsym_2)$ are distributed according to a  bivariate Rayleigh fading distribution with pdf 
\begin{align}\nonumber
f_{\Rsym_1,}&_{\Rsym_2}(\rsym_1,\rsym_2)=\\\nonumber
&\frac{4\rsym_1\rsym_2}{\srxq\sryq(1-\fcor)}\exp\bigg(-\frac{1}{1-\fcor}\bigg(\frac{\rsym_1^2}{\srxq}+\frac{\rsym_2^2}{\sryq}\bigg)\bigg)\\\label{eq:fading-pdf}
&\quad\times I_{0}\Bigg(\frac{2}{1-\fcor}\sqrt{\fcor\frac{\rsym_1^2}{\srxq}\frac{\rsym_2^2}{\sryq}}\Bigg),\quad \rsym_1,\rsym_2 \geq 0
\end{align}
where $I_0(x)=\frac{1}{\pi}\int_{0}^{\pi}e^{x\cos(\phi)}\mathrm{d}\phi$ denotes the zeroth-order modified Bessel function of the first kind  
\cite[10.25.2, 10.32.1]{Olver2010}.   
The parameters $\srxq$ and $\sryq$ in \eqref{eq:fading-pdf} are 
\begin{align}
\srxq=\E\big[\Rsym_1^2\big],&\qquad \sryq=\E\big[\Rsym_2^2\big]
\end{align}
denoting (with a slight abuse of common notation) the second moments of $\Rsym_1$ and $\Rsym_2$. 
Furthermore, $\fcor$, for $0 \leq \fcor < 1$, denotes the power correlation coefficient, i.e., we have
\begin{align}
\mathrm{cor}(\Rsym_1^2,\Rsym_2^2)=\fcor 
\end{align}
which is the Pearson correlation coefficient between $\Rsym_1^2$ and $\Rsym_2^2$. For later reference, we provide the marginal pdfs $f_{\Rsym_1}$ and $f_{\Rsym_2}$ of $\Rsym_1$ and $\Rsym_2$ 
\begin{alignat}{2}\label{eq:pdf-r1}
f_{\Rsym_1}(\rsym_1) &=\frac{2\,\rsym_1}{\sigma_{\Rsym_1}^2}\exp\bigg(-\frac{\rsym_1^2}{\sigma_{\Rsym_1}^2}\bigg),\quad &&\rsym_1 \geq 0,\\\label{eq:pdf-r2} 
f_{\Rsym_2}(\rsym_2) &=\frac{2\rsym_2}{\sryq}\exp\bigg(-\frac{\rsym_2^2}{\sryq}\bigg), &&\rsym_2 \geq 0
\end{alignat}
as well as further moments
\begin{gather}\nonumber
\E\big[\Rsym_1\big]=\sqrt{\frac{\pi}{4}\srxq},\quad
\E\big[\Rsym_2\big]=\sqrt{\frac{\pi}{4}\sryq}, \\[1ex]\label{eq:rayleigh-variance}
\mathrm{var}\big[\Rsym_1\big]=\Big(1-\frac{\pi}{4}\Big)\srxq,\quad
\mathrm{var}\big[\Rsym_2\big]=\Big(1-\frac{\pi}{4}\Big)\sryq, \\[1ex]\label{eq:rayleigh-co-variance}
\mathrm{cov}\big[\Rsym_1,\Rsym_2\big]\!=\!\srx\sry\bigg(E\Big(\sqrt{\fcor}\Big)\!-\!\frac{1}{2}\big(1-\fcor\big)K\Big(\sqrt{\fcor}\Big)-\frac{\pi}{4}\bigg)
\end{gather}
where $\mathrm{cov}[\cdot,\cdot]$ and $\mathrm{var}[\cdot]$  denote (co)variance and 
\begin{align}\label{eq:elliptic-k}
K(z)&=\int_{0}^{\frac{\pi}{2}}\big(1-z^2\sin^2(t)\big)^{-\frac{1}{2}}\,\mathrm{d}t\\\label{eq:elliptic-e}
E(z)&=\int_{0}^{\frac{\pi}{2}}\big(1-z^2\sin^2(t)\big)^{\frac{1}{2}}\,\mathrm{d}t
\end{align}
are the complete elliptic integrals of the first and second kind \cite[19.2.4,\,19.2.5,\,19.2.8]{Olver2010}. Moreover, with the marginals \eqref{eq:pdf-r1} and \eqref{eq:pdf-r2} and basic transformations we obtain the complementary cumulative distribution functions of $\Rsym_1^2$ and $\Rsym_2^2$ as 
\begin{align*}
\bar{F}_{\Rsym_1^2}(\rsym_1) &=\exp\bigg(\!-\frac{\rsym_1}{\srxq}\bigg),\quad\rsym_1 \geq 0,\\
\bar{F}_{\Rsym_2^2}(\rsym_2)&=\exp\bigg(\!-\frac{\rsym_2}{\sryq}\bigg), \quad\rsym_2 \geq 0. 
\end{align*}
Therefore, for the bivariate Rayleigh distribution, the condition \eqref{eq:condition-stochastic-degradedness} on stochastic degradedness is equivalent to
\begin{align}\label{eq:cond-degraded-rayleigh}
\frac{\sigma_{N_2}^2}{\sigma_{N_1}^2} \;\leq\; \frac{\srxq}{\sryq}. 
\end{align}
%

\section{Achievable Rates for Gaussian Input}
\label{sec:achievable-rate}
Given (\ref{Eq:Achievable-Rate}), the main goal is to find the maximum of its right-hand side w.\,r.\,t.\ the distribution $P_X$ of the random variable $X$. However, this is a difficult optimization problem, so we provide instead an achievable rate for a Gaussian input $X$. Subsequently, we evaluate \eqref{Eq:Achievable-Rate-a}--\eqref{Eq:Achievable-Rate-c} for $X$ being a zero-mean Gaussian random variable with positive variance $P$, where $X$ is independent of $(\Rsym_1,\Rsym_2,N_1,N_2)$.

\subsection{Evaluation of Eq.~\eqref{Eq:Achievable-Rate-a}}\label{sec:eval-part-a}

\begin{proposition}\label{Prop:Part-A} 
Under the assumptions above, we have 
\begin{align}\nonumber
\E_{\Rsym_1,\Rsym_2}\big[h&(\Rsym_1X+N_1|\Rsym_2X+N_2)\big] -h(N_1)\\\label{eq:integral-prop-part-a}
 &= \frac{1}{2}\int_{0}^{\infty} \log_2(1+\rsym)f_\Rsym(\rsym)\,\mathrm{d}\rsym
\end{align}
where 
\begin{gather}\nonumber
f_\Rsym(\rsym)=\sigma_1^2\sigma_2^2\exp\bigg(\frac{\sigma_2^2}{2P(1-\fcor)}\bigg)\exp\bigg(-\frac{\sigma_1^2 \rsym +\sqrt{A(\rsym)}}{2P(1-\fcor)}\bigg)\\\label{eq:density-rv-r}
\quad\;\;\;\times\bigg(\frac{1}{2P\sqrt{A(\rsym)}}\!+\!\frac{\sigma_1^2 \rsym \!+\! \sigma_2^2}{2PA(\rsym)}\!+\!(1\!-\!\fcor)\frac{\sigma_1^2 \rsym + \sigma_2^2}{A(\rsym)^{\frac{3}{2}}}\bigg)
\intertext{with}\label{eq:density-rv-r-polynom}
A(\rsym) = (\sigma_1^2 \rsym)^2 + (2-4\fcor) \sigma_1^2 \sigma_2^2 \rsym + (\sigma_2^2)^2,\\[1ex]\label{eq:params-rv-r}
\sigma_1^2=\frac{\sigma_{N_1}^2}{\srxq},\quad \sigma_2^2=\frac{\sigma_{N_2}^2}{\sryq}.
\end{gather}
\end{proposition}

The derivation of Proposition~\ref{Prop:Part-A} is given in Appendix~\ref{app:proof-prop:Part-A}. The representation in Proposition~\ref{Prop:Part-A} as a one-dimensional integral is particularly convenient for numerical evaluations and is used in Section~\ref{sec:results}. 
%

\subsection{Evaluation of Eq.~\eqref{Eq:Achievable-Rate-b}}\label{sec:eval-part-b}

First, we rewrite \eqref{Eq:Achievable-Rate-b} as 
\begin{align*}
\E_{X}\big[h(\Rsym_1X+N_1|\Rsym_2)\big] = \E_{X}\big[h(\Rsym_1X+N_1,\Rsym_2)\big] - h(\Rsym_2).
\end{align*}
Using the marginal pdf \eqref{eq:pdf-r2} of $\Rsym_2$, we obtain
\begin{align}\nonumber
h(\Rsym_2) = & -\int_{0}^\infty f_{\Rsym_2}(\rsym_2)\log_2\big(f_{\Rsym_2}(\rsym_2)\big)\,\mathrm{d}\rsym_2\\\nonumber
= & -2\log_2\bigg(\frac{2}{\sry}\bigg)\int_{0}^{\infty}u\exp\big(-u^2\big)\,\mathrm{d}u\\\nonumber
 & \quad -2\int_{0}^{\infty}u\exp\big(-u^2\big)\log_2(u)\,\mathrm{d}u\\\nonumber
& \quad + \frac{2}{\ln(2)}\int_{0}^{\infty}u^3\exp\big(-u^2\big)\,\mathrm{d}u\\\label{eq:part-b-diff-h-r2}
=&\,\frac{1}{\ln(2)}\bigg(1+\frac{\gamma}{2}\bigg)+\frac{1}{2}\log_2\bigg(\frac{\sryq}{4}\bigg)
\end{align}
using the substitution $u=\rsym_2/\srx$ and the integral relations	 $\int_{0}^{\infty}u\exp(-u^2)\,\mathrm{d}u=\int_{0}^{\infty}u^3\exp(-u^2)\,\mathrm{d}u=1/2$ and $\int_{0}^{\infty}u\exp(-u^2)\ln(u)\,\mathrm{d}u = -\gamma/4$, where $\gamma = 0.577216\ldots$ denotes Euler's constant \cite[5.2.3]{Olver2010}. 

The evaluation of $\E_{X}\big[h(\Rsym_1X+N_1,\Rsym_2)\big]$ requires the following calculations.  
Let  $Y_1(x)=x\Rsym_1+N_1$. Then the joint pdf of $\big(Y_1(x),\Rsym_2\big)$ is given for $x>0$ by the convolution integral  
\begin{gather}\label{eq:pdf-convolution-integral}
f_{Y_1(x),\Rsym_2}(y_1,\rsym_2)\!=\!\int_{0}^{\infty}\!\!\frac{1}{x}f_{\Rsym_1,\Rsym_2}\bigg(\frac{t}{x},\rsym_2\bigg)f_{N_1}(y_1\!-\!t)\mathrm{d}t 
\end{gather}
for $-\infty < y_1 < \infty$ and $\rsym_2\geq 0$. Furthermore, we have 
\begin{align}\label{eq:h-for-fixed-x-part-b}
&h(x\Rsym_1+N_1,\Rsym_2) =\\\nonumber
& -\hspace*{-1.25em}\int\limits_{y_1=-\infty}^{\infty}\int\limits_{\rsym_2=0}^{\infty} f_{Y_1(x),\Rsym_2}(y_1,\rsym_2)\log_2\Big(f_{Y_1(x),\Rsym_2}(y_1,\rsym_2)\Big)\,\mathrm{d}y_1\mathrm{d}\rsym_2.
\end{align}
Due to symmetry, we obtain
\begin{align}\label{eq:expectation-h-part-b}
&\E_{X}\big[h(\Rsym_1X + N_1,\Rsym_2)\big]\nonumber\\
&\qquad\qquad = 2\int_{0}^{\infty} h(x\Rsym_1+N_1,\Rsym_2)f_X(x)\,\mathrm{d}x. 
\end{align}
As we can evaluate the convolution integral in \eqref{eq:pdf-convolution-integral} numerically, we rely on numerical calculations also for \eqref{eq:h-for-fixed-x-part-b} and \eqref{eq:expectation-h-part-b}.

An upper bound of $\E_{X}\big[h(\Rsym_1X+N_1|\Rsym_2)\big]$, for which the numerical evaluation is much easier is the following.  
\begin{proposition}\label{prop:part-b-bound} Under the assumptions above, we have
\begin{align}\nonumber
\E_{X}&\big[h(\Rsym_1X+N_1|\Rsym_2)\big]\\\nonumber 
 \leq\, &\frac{1}{2}\log_2\big(2(\pi\mathrm{e})^2\srxq\bcor\,P\big)
+\frac{\pi}{2\ln(2)}\,\mathrm{erfi}\bigg(\sqrt{\frac{\sxtd}{2\,\bcor\, P}}\bigg)+1\\\label{eq:eval-bound-b} 
&-\frac{1}{\ln(2)}\bigg(\frac{\sxtd}{2\,\bcor\,P}\, {}_2F{}_2\big(1,1;\tfrac{3}{2},2;\frac{\sxtd}{2\,\bcor\,P}\big)  +1+\gamma\bigg) 
\end{align}
with the parameters 
\begin{align}\label{eq:part-b-bound-params-a}
\sxtd \!=\! \Big(1\!-\!\frac{\pi}{4}\Big)\frac{\sigma_{N_1}^2}{\srxq},
\quad\bcor \!=\!\Big(1\!-\!\frac{\pi}{4}\Big)^2\Big(1\!-\!\mathrm{cor}(\Rsym_1,\Rsym_2)^2\Big)
\end{align}
where 
\begin{align}
&\mathrm{cor}[\Rsym_1,\Rsym_2]=\frac{\mathrm{cov}[\Rsym_1,\Rsym_2]}{\sqrt{\mathrm{var}[\Rsym_1]\mathrm{var}[\Rsym_2]}}\\
 &\; =\Big(1-\frac{\pi}{4}\Big)^{-1}\bigg(E\Big(\sqrt{\fcor}\Big)-\frac{1}{2}\big(1-\fcor\big)K\Big(\sqrt{\fcor}\Big)-\frac{\pi}{4}\bigg)  
\end{align}
with $K(\cdot)$ and $E(\cdot)$ the elliptic integrals given in \eqref{eq:elliptic-k} and \eqref{eq:elliptic-e}, respectively. Moreover,  $\mathrm{erfi}(y)=-\imath\,\mathrm{erf}(\imath y)=-\frac{2\imath}{\sqrt{\pi}}\int_{0}^{\imath y}\exp\big(-t^2\big)\,\mathrm{d}t$  denotes the imaginary error function \cite[7.2.1]{Olver2010} and ${}_pF{}_q(a_1,\ldots,a_p;b_1,\ldots,b_q;z)$ denotes the generalized hypergeometric function \cite[16.2.1]{Olver2010}.
\end{proposition}

The proof of Proposition~\ref{prop:part-b-bound} is given in Appendix~\ref{app:prop:part-b-bound}.

\subsection{Evaluation of Eq.~\eqref{Eq:Achievable-Rate-c}}\label{sec:eval-part-c}

\begin{proposition}\label{Prop:Part-C} Under the assumptions above, we have
\begin{align}\nonumber
\E_{\Rsym_1}\big[&h(\Rsym_1X+N_1)] -h(N_1)\\\label{eq:Prop:Part-C}
 &= -\frac{1}{2\ln(2)}\exp\bigg(\frac{1}{P}\bigg)\mathrm{Ei}\bigg(\frac{1}{P}\bigg),
\end{align}
where $\mathrm{Ei}(z)=-\int_{-z}^\infty\frac{\exp(-t)}{t}\,\mathrm{d}t$ denotes the exponential integral function \cite[6.2.5]{Olver2010}. 
\end{proposition}

The derivation of Proposition~\ref{Prop:Part-C} is given in Appendix~\ref{app:proof-prop:Part-C}.

\section{Numerical Results and Discussions}
\label{sec:results}
We next evaluate the results of Section~\ref{sec:achievable-rate} numerically for interesting parameter regimes.  
To simplify notation, we denote the sum of \eqref{Eq:Achievable-Rate-a} and \eqref{Eq:Achievable-Rate-b} by 
$\Ralpha$ and the sum of \eqref{Eq:Achievable-Rate-a} and the upper bound \eqref{eq:eval-bound-b} of \eqref{Eq:Achievable-Rate-b} by $\RalphaUB$, respectively. Furthermore, we denote \eqref{Eq:Achievable-Rate-c} by $\Rbeta$. With this notation, we have for the achievable rate in Section~\ref{sec:achievable-rate}
\begin{align}
R \leq \,\min\big\{\Ralpha,\Rbeta\big\} \leq \,\min\big\{\RalphaUB,\Rbeta\big\}. 
\end{align}

Based on the representation in \eqref{eq:integral-prop-part-a}--\eqref{eq:density-rv-r-polynom} as one-dimensional integral, we numerically evaluate \eqref{Eq:Achievable-Rate-a} with \textsc{Mathematica}. Similarly, the upper bound in \eqref{eq:eval-bound-b} is numerically evaluated, and the same applies to 
\eqref{Eq:Achievable-Rate-c} using \eqref{eq:Prop:Part-C}. 
However, the numerical evaluation of \eqref{Eq:Achievable-Rate-b} is more involved. First, we  numerically calculate the convolution integral in \eqref{eq:pdf-convolution-integral} on a sufficiently-dense grid for the variables $y_1$ and $\rsym_2$. Then, we numerically calculate the differential entropy $h(x\Rsym_1+N_1,\Rsym_2)$ using \eqref{eq:h-for-fixed-x-part-b} based on an interpolated version of the density 
$f_{Y_1(x),\Rsym_2}(y_1,\rsym_2)$. Repeating these calculations for a sufficiently-dense set of values $x$, we  numerically calculate $\E_{X}\big[h(\Rsym_1X+N_1,\Rsym_2)\big]$ using \eqref{eq:expectation-h-part-b} and an interpolated version of the function $x\!\mapsto\! h(x\Rsym_1\!+\!N_1,\Rsym_2).$ Combining with \eqref{eq:part-b-diff-h-r2}, we finally obtain \eqref{Eq:Achievable-Rate-b}.

We consider the case of a stochastically-degraded secure ISAC channel, i.\,e., we assume the chosen parameter values satisfy   inequality \eqref{eq:cond-degraded-rayleigh}. Moreover, we assume that $\sigma_{N_1}^2\!>\!\sigma_{N_2}^2$, which is the interesting regime where the corresponding wiretap channel does not allow secure communication. Table~\ref{tab:parameter-sets} summarizes parameter sets satisfying these conditions for which we subsequently present and discuss numerical results.  

In Figs.\,\ref{fig:bounds-rho-01}--\ref{fig:bounds-rho-09}, given in Appendix~\ref{app:figures}, we illustrate the results for $\Ralpha$, $\RalphaUB$, and $\Rbeta$ as a function of the transmit power $P$ for different values of the power correlation coefficient $\fcor$. For the matrix of subfigures in each figure, the parameter $\srxq$ is modified from top to bottom and the parameter $\sigma_{N_2}^2$ from left to right, respectively, whereas $\sigma_{N_1}^2=1$ is fixed. The parameter  $\sryq$ is modified within a subfigure. 

Our conclusions for a degraded secure ISAC channel with correlated Rayleigh fading for the parameter ranges given in Table~\ref{tab:parameter-sets} are discussed next.

From \eqref{eq:Prop:Part-C}, we observe that $\Rbeta$ is only a function of the transmit power $P$ such that the curves of $\Rbeta$ are the same in all diagrams. From \eqref{eq:integral-prop-part-a}--\eqref{eq:density-rv-r-polynom}, we observe that \eqref{Eq:Achievable-Rate-a} as a summand of $\Ralpha$ and $\RalphaUB$ is a function of  
$\fcor$, $P$, and the parameter ratios ${\sigma_{N_1}^2}/{\srxq}$ and ${\sigma_{N_2}^2}/{\sryq}$. Similarly, the upper bound \eqref{eq:eval-bound-b} as a summand of $\RalphaUB$, is a function of $\fcor$, $P$, $\srxq$, and $\sigma_{N_1}^2$ and it does not depend on $\sryq$ and $\sigma_{N_2}^2$. Although not explicit from the derived equations, the numerical results also show that \eqref{Eq:Achievable-Rate-b} as a summand of $\Ralpha$ does not depend on $\sryq$ and $\sigma_{N_2}^2$. Therefore, the curves of $\Ralpha$ and $\RalphaUB$ in Figs.\,\ref{fig:bounds-rho-01}--\ref{fig:bounds-rho-09} within one subfigure and between two subfigures in the same row differ only due to the summand \eqref{Eq:Achievable-Rate-a}.

The results show that $\RalphaUB$ and $\Ralpha$ curves behave highly similarly with a small constant gap. Thus, for most of the parameter constellations  the much-easier-to-calculate $\RalphaUB$, instead of $\Ralpha$, can be used to interpret the results.

\begin{table}[tb]
\centering
	\begin{tabular}{|c|c|}\hline
	\multicolumn{2}{|c|}{$\fcor\in\{0.01,0.50,0.90\}$\rule{0ex}{2.5ex}}\\[1ex]\hline
		$\sigma_{N_1}^2=1$ & $\sigma_{N_2}^2\in\{0.10,0.50\}$ \rule{0ex}{2.5ex}\\[1ex]\hline
		$\srxq\in\{0.10,0.50,1.00\}$ & $\sryq\in\left\{{\srxq}/{\sigma_{N_2}^2}\,,\,{\srxq}/{10\sigma_{N_2}^2}\right\}$\rule{0ex}{3.75ex}\\[2ex]\hline
	\end{tabular}\\[1.5ex]
\caption{Parameter sets for numerical calculations}
\label{tab:parameter-sets}
\vspace{-0.55cm}
\end{table}

Furthermore, we observe the following monotonicities: $\Ralpha$ increases for increasing parameters $\srxq$ or $\sigma_{N_2}^2$ and for decreasing parameters $\sryq$ or $\fcor$. The interesting regime where the channel capacity is approached is when $\Rbeta$ determines the right-hand side of \eqref{Eq:Achievable-Rate}. The range for the power $P$ where 
$\Rbeta$ determines \eqref{Eq:Achievable-Rate}, increases with increasing $\srxq$ or $\sigma_{N_2}^2$ and decreasing $\sryq$ or $\fcor$. For low correlation $\fcor$, this range stretches over all considered power values for almost all parameter constellations, whereas for highly correlated fading coefficients it shrinks to low power values. Thus, in the low power regime channel capacity is always approached irrespective of the values of the remaining parameters.

\section*{Acknowledgment}
This work has been supported by the German Federal Ministry of Education and Research (BMBF) through the research hub \emph{6G-life} under grant 16KISK001K, the German Research Foundation (DFG)  as part of Germany’s Excellence Strategy -- EXC 2050/1 - Project ID 390696704 - Cluster of Excellence \emph{CeTI}, the U.S. National Science Foundation (NSF) under grants CCF 1955401 and 2148400, the U.S. Department of Transportation under grant 69A3552348327 for the CARMEN+ University Transportation Center, the ZENITH Research and Leadership Career Development Fund, Chalmers Transport Area of Advance, and the ELLIIT funding endowed by the Swedish government.

\bibliographystyle{IEEEtran}
\bibliography{referencesGlobecom2024}

\begin{thebibliography}{10}
\providecommand{\url}[1]{#1}
\csname url@samestyle\endcsname
\providecommand{\newblock}{\relax}
\providecommand{\bibinfo}[2]{#2}
\providecommand{\BIBentrySTDinterwordspacing}{\spaceskip=0pt\relax}
\providecommand{\BIBentryALTinterwordstretchfactor}{4}
\providecommand{\BIBentryALTinterwordspacing}{\spaceskip=\fontdimen2\font plus
\BIBentryALTinterwordstretchfactor\fontdimen3\font minus
  \fontdimen4\font\relax}
\providecommand{\BIBforeignlanguage}[2]{{%
\expandafter\ifx\csname l@#1\endcsname\relax
\typeout{** WARNING: IEEEtran.bst: No hyphenation pattern has been}%
\typeout{** loaded for the language `#1'. Using the pattern for}%
\typeout{** the default language instead.}%
\else
\language=\csname l@#1\endcsname
\fi
#2}}
\providecommand{\BIBdecl}{\relax}
\BIBdecl

\bibitem{NokiaGuysJCASTutorial}
T.~Wild, V.~Braun, and H.~Viswanathan, ``Joint design of communication and
  sensing for beyond {5G} and {6G} systems,'' \emph{{IEEE} {A}ccess}, vol.~9,
  pp. 30\,845--30\,857, Feb. 2021.

\bibitem{JCASwithSecurityTutorial}
Z.~Wei, F.~Liu, C.~Masouros, N.~Su, and A.~P. Petropulu, ``Towards
  multi-functional {6G} wireless networks: {I}ntegrating sensing, communication
  and security,'' July 2021, [Online]. Available: arxiv.org/abs/2107.07735.

\bibitem{VDEJCASPositionPaper}
G.~Fettweis \emph{et~al.}, ``Joint communications \& sensing - {C}ommon
  radio-communications and sensor technology,'' \emph{{VDE} Positionspapier},
  July 2021.

\bibitem{Zhang_2011}
W.~Zhang, S.~Vedantam, and U.~Mitra, ``Joint transmission and state estimation:
  {A} constrained channel coding approach,'' \emph{{IEEE} Trans. Inf. Theory},
  vol.~57, no.~10, pp. 7084--7095, Oct. 2011.

\bibitem{AcademicsJCASTutorial}
H.~Wymeersch \emph{et~al.}, ``Integration of communication and sensing in {6G}:
  {A} joint industrial and academic perspective,'' in \emph{{IEEE} Annu. Int.
  Symp. Pers., Indoor Mobile Radio Commun.}, Helsinki, Finland, Sep. 2021, pp.
  1--7.

\bibitem{MassiveMIMOforJCAS}
S.~Buzzi, C.~D'Andrea, and M.~Lops, ``Using {Massive} {MIMO} arrays for joint
  communication and sensing,'' in \emph{{A}silomar Conf. Signals, Syst.,
  Comput.}, Pacific Grove, CA, Nov. 2019, pp. 5--9.

\bibitem{MariMichelleGJournalBroadcast}
M.~Ahmadipour, M.~Kobayashi, M.~Wigger, and G.~Caire, ``An
  information-theoretic approach to joint sensing and communication,''
  \emph{{IEEE} Trans. Inf. Theory}, May 2022.

\bibitem{MariMACISIT}
M.~Kobayashi, H.~Hamad, G.~Kramer, and G.~Caire, ``Joint state sensing and
  communication over memoryless multiple access channels,'' in \emph{{IEEE}
  Int. Symp. Inf. Theory}, Paris, France, July 2019, pp. 270--274.

\bibitem{MichelleCollaborative}
M.~Ahmadipour and M.~Wigger, ``An information-theoretic approach to
  collaborative integrated sensing and communication for two-transmitter
  systems,'' \emph{{IEEE} J. Sel. Areas Inf. Theory}, vol.~4, pp. 112--127,
  June 2023.

\bibitem{TrumanISIT2024}
T.~Welling, O.~G{\"u}nl{\"u}, and A.~Yener, ``Transmitter actions for secure
  integrated sensing and communication,'' in \emph{{IEEE} Int. Symp. Inf.
  Theory}, Athens, Greece, July 2024, to appear.

\bibitem{oursecureISACJSAIT}
O.~G{\"u}nl{\"u}, M.~Bloch, R.~Schaefer, and A.~Yener, ``Secure integrated
  sensing and communication,'' \emph{{IEEE} J. Selected in Inf. Theory},
  vol.~4, pp. 40--53, May 2023.

\bibitem{OurJCandSConferenceISAC}
O.~Günlü, M.~Bloch, R.~F. Schaefer, and A.~Yener, ``Secure integrated sensing
  and communication for binary input additive white {Gaussian} noise
  channels,'' in \emph{{IEEE} Int. Symp. Joint Commun. \& Sensing}, Seefeld,
  Austria, Mar. 2023, pp. 1--6.

\bibitem{MichelleSecureISACwithSholomo}
M.~Ahmadipour, M.~Wigger, and S.~Shamai, ``Integrated communication and
  receiver sensing with security constraints on message and state,'' in
  \emph{{IEEE} Int. Symp. Inf. Theory}, Taipei, Taiwan, June 2023, pp.
  2738--2743.

\bibitem{AhlswedeCaiWTCwithFeedback}
R.~Ahlswede and N.~Cai, ``Transmission, identification and common randomness
  capacities for wire-tape channels with secure feedback from the decoder,''
  \emph{Electron. Notes Discrete Math.}, vol.~21, pp. 155--159, Aug. 2005.

\bibitem{AsafCohenWTCwithFeedback}
A.~Cohen and A.~Cohen, ``Wiretap channel with causal state information and
  secure rate-limited feedback,'' \emph{{IEEE} Trans. Commun.}, vol.~64, no.~3,
  pp. 1192--1203, Mar. 2016.

\bibitem{OurJSAITTutorial}
M.~Bloch, O.~G{\"u}nl{\"u}, A.~Yener, F.~Oggier, H.~V. Poor, L.~Sankar, and
  R.~F. Schaefer, ``An overview of information-theoretic security and privacy:
  {M}etrics, limits and applications,'' \emph{{IEEE} J. Sel. Areas Inf.
  Theory}, vol.~2, no.~1, pp. 5--22, Mar. 2021.

\bibitem{HanVinckWTCwithFeedback}
B.~Dai, A.~J.~H. Vinck, Y.~Luo, and Z.~Zhuang, ``Capacity region of
  non-degraded wiretap channel with noiseless feedback,'' in \emph{{IEEE} Int.
  Symp. Inf. Theory}, Cambridge, MA, July 2012, pp. 244--248.

\bibitem{he-yener-fbsecrecy}
X.~{He} and A.~{Yener}, ``The role of feedback in two-way secure
  communications,'' \emph{{IEEE} Trans. Inf. Theory}, vol.~59, no.~12, pp.
  8115--8130, Dec. 2013.

\bibitem{GermanWTCwithGeneralizedFeedback}
G.~Bassi, P.~Piantanida, and S.~Shamai, ``The wiretap channel with generalized
  feedback: {S}ecure communication and key generation,'' \emph{{IEEE} Trans.
  Inf. Theory}, vol.~65, no.~4, pp. 2213--2233, Apr. 2019.

\bibitem{TseFadingBC}
D.~Tse and R.~Yates, ``Fading broadcast channels with state information at the
  receivers,'' \emph{{IEEE} Trans. Inf. Theory}, vol.~58, no.~6, pp.
  3453--3471, June 2012.

\bibitem{OurPUFTutorial}
O.~G{\"u}nl{\"u} and R.~F. Schaefer, ``An optimality summary: {Secret} key
  agreement with physical unclonable functions,'' \emph{Entropy}, vol.~23,
  no.~1, 2020.

\bibitem{ISACMACChinesegroup}
Y.~Liu \emph{et~al.}, ``Generalized modeling and fundamental limits for
  multiple-access integrated sensing and communication systems,'' May 2022,
  [Online]. Available: arxiv.org/abs/2205.05328.

\bibitem{Elgamalbook}
A.~E. Gamal and Y.-H. Kim, \emph{Network {I}nformation {T}heory}.\hskip 1em
  plus 0.5em minus 0.4em\relax Cambridge, {U.K.}: Cambridge {U}niversity
  {P}ress, 2011.

\bibitem{PinHsunGaussianWiretap}
P.-H. Lin and E.~Jorswieck, ``On the fast fading {Gaussian} wiretap channel
  with statistical channel state information at the transmitter,'' \emph{{IEEE}
  Trans. Inf. Forensics Security}, vol.~11, no.~1, pp. 46--58, Jan. 2016.

\bibitem{Olver2010}
F.~W.~J. Olver, D.~W. Lozier, R.~F. Boisvert, and C.~W. Clark, Eds., \emph{NIST
  Handbook of Mathematical Functions}.\hskip 1em plus 0.5em minus 0.4em\relax
  Cambridge, UK: Cambridge University Press, 2010.

\bibitem{Papoulis2002}
A.~Papoulis and S.~U. Pillai, \emph{Probability, Random Variables, and
  Stochastic Processes}, 4th~ed.\hskip 1em plus 0.5em minus 0.4em\relax Boston,
  MA, US: McGraw-Hill, 2002.

\bibitem{Prudnikov1986a}
A.~P. Prudnikov, Y.~A. Brychov, and O.~I. Marichev, \emph{Integrals and Series,
  Volume 2: Special Functions}.\hskip 1em plus 0.5em minus 0.4em\relax New
  York, NY, USA: Gordon and Breach Science, 1986.

\bibitem{Prudnikov1986}
------, \emph{Integrals and Series, Volume 1: Elementary Functions}.\hskip 1em
  plus 0.5em minus 0.4em\relax New York, NY, USA: Gordon and Breach, 1986.

\end{thebibliography}


\appendices

\section{Proof of Proposition~\ref{Prop:Part-A}}\label{app:proof-prop:Part-A}

\begin{IEEEproof} Let us fix $\Rsym_1=\rsym_1$ and $\Rsym_2=\rsym_2$ for some $\rsym_1,\rsym_2 \geq 0$. Then $(\rsym_1X+N_1,\rsym_2X+N_2)$ is jointly Gaussian with zero mean and covariance matrix
\begin{align*}
\begin{pmatrix}
\rsym_1^2P+\sigma_{N_1}^2 & \rsym_1\rsym_2 P\\
\rsym_1\rsym_2 P & \rsym_2^2P+\sigma_{N_2}^2
\end{pmatrix}
\end{align*}
since $X$, $N_1$, and $N_2$ are independent Gaussian random variables with positive variances $P$, $\sigma_{N_1}^2$, and $\sigma_{N_2}^2$. 
For the conditional differential entropy $h\big(\rsym_1X+N_1|\rsym_2X+N_2\big)$, we therefore obtain
\begin{align}\nonumber
h&\big(\rsym_1X+N_1\,|\,\rsym_2X+N_2\big)\\\nonumber
 &= h\big(\rsym_1X+N_1,\rsym_2X+N_2\big)-h\big(\rsym_2X+N_2\big)\\\label{eq:part-a-ratio-represenation}
& = \frac{1}{2} \log_2\big(2\pi\mathrm{e}\sigma_{N_1}^2\big)+ \frac{1}{2} \log_2\Bigg(1+\frac{\rsym_1^2/\sigma_{N_1}^2}{\rsym_2^2/\sigma_{N_2}^2+\nicefrac{1}{P}}\Bigg).
\end{align}
Using \eqref{eq:part-a-ratio-represenation}, we can write  
\begin{align*}
\E_{\Rsym_1,\Rsym_2}&\big[h(\Rsym_1X+N_1|\Rsym_2X+N_2)\big]\\
 &= h(N_1)+\frac{1}{2}\,\E_{\Rsym}\big[\log_2\big(1+\Rsym\big)\big]
\end{align*}
where the random variable $\Rsym$ is given by
\begin{align*}
\Rsym=\frac{T_1}{T_2+\nicefrac{1}{P}} 
\end{align*}
with $(T_1,T_2)=\big(\Rsym_1^2/\sigma_{N_1}^2,\Rsym_2^2/\sigma_{N_2}^2\big)$.
Since $\Rsym$ is a ratio of random variables, the pdf $f_\Rsym$ of $\Rsym$ has the following integral representation 
\cite[Eq.\,6.60]{Papoulis2002} 
\begin{align}\label{eq:pdf-rv-r}
f_\Rsym(\rsym) = \int_{\nicefrac{1}{P}}^\infty u f_{T_1,T_2}\big(\rsym\cdot u,u-\nicefrac{1}{P}\big)\,\mathrm{d}u,\quad \rsym\geq 0
\end{align}
where $f_{T_1,T_2}$ is the joint pdf of $(T_1,T_2)$ given by
\begin{align}\nonumber 
f_{T_1,T_2}(t_1,&t_2)=
\,\frac{\sigma_1^2\sigma_2^2}{(1-\fcor)}\exp\bigg(-\frac{1}{1-\fcor}\Big(\sigma_1^2t_1+\sigma_2^2t_2\Big)\bigg)\\\label{eq:scaled-fading-pdf}
&\times I_{0}\bigg(\frac{2}{1-\fcor}\sqrt{\fcor\,\sigma_1^2t_1\,\sigma_2^2t_2}\bigg),\quad t_1,t_2 \geq 0
\end{align}
where $\sigma_1^2$ and $\sigma_2^2$ are as specified in \eqref{eq:params-rv-r}. 

We substitute $\tilde{u}=(u - \nicefrac{1}{P})$ in \eqref{eq:pdf-rv-r}, plug it in \eqref{eq:scaled-fading-pdf}, and then subsitute $v=\sqrt{\tilde{u}(\tilde{u}+\nicefrac{1}{P})\,}$ to obtain 
\begin{align*}
&f_\Rsym(\rsym)=\frac{\sigma_1^2\sigma_2^2}{1-\fcor}\exp\bigg(\frac{\sigma_2^2}{2P(1-\fcor)}\bigg)\exp\bigg(-\frac{\sigma_1^2\rsym}{2P(1-\fcor)}\bigg)\\
&\;\times\int_{0}^{\infty}\bigg[\frac{\theta v}{\sqrt{v^2+\theta^2}}+v\bigg]\exp\Big(-\alpha\sqrt{v^2+\theta^2}\Big)I_{0}\big(\beta v\big)\,\mathrm{d}v 
\end{align*}%
where 
\begin{gather*}
\alpha=\frac{\sigma_1^2\rsym+\sigma_2^2}{1-\fcor},\quad \beta=2\frac{\sqrt{\fcor\sigma_1^2\sigma_2^2\,\rsym}}{1-\fcor},\quad \theta =\frac{1}{2P}. 
\end{gather*}
Using this representation, we can directly apply \cite[Sec.\,2.5.6.10]{Prudnikov1986a} and \cite[Sec.\,2.5.6.13]{Prudnikov1986a}. After collecting terms we finally obtain the form of the density $f_\Rsym$ given in \eqref{eq:density-rv-r}, which completes the proof. 
\end{IEEEproof}

\section{Proof of Proposition~\ref{prop:part-b-bound}}\label{app:prop:part-b-bound}
\begin{IEEEproof}
Fix $X=x$ for some $x\in\R$. Then, the differential entropy $h\big(x\Rsym_1+N_1,\Rsym_2\big)$ is bounded by
\begin{align*}
h\big(x\Rsym_1+N_1,\Rsym_2\big) \leq \frac{1}{2}\log_2\bigg(&(2\pi\mathrm{e})^2\Big(\mathrm{var}\big[x\Rsym_1+N_1\big]\mathrm{var}\big[\Rsym_2\big]\\
&-\big(\mathrm{cov}\big[x\Rsym_1+N_1,\Rsym_2\big)\big]^2\Big)\bigg)
\end{align*}
as a result of the differential entropy maximizing property of the Gaussian distribution with a given covariance matrix.
Since $\Rsym_1,\Rsym_2$, and $N_1$ are independent, we have 
\begin{align*}
\mathrm{var}\big[x\Rsym_1+N_1\big] &= x^2\mathrm{var}[\Rsym_1]+\mathrm{var}[N_1],\\
\mathrm{cov}\big[x\Rsym_1+N_1,\Rsym_2\big] &=x\,\mathrm{cov}\big[\Rsym_1,\Rsym_2\big].
\end{align*}
With the variance $\mathrm{var}[\Rsym_1]$ and covariance $\mathrm{cov}\big[\Rsym_1,\Rsym_2\big]$ given in \eqref{eq:rayleigh-variance} and  \eqref{eq:rayleigh-co-variance}, we obtain 
\begin{align*}
h\big(x\Rsym_1+N_1,\Rsym_2\big) \leq &\;a+\frac{1}{2}\log_2\big(\bcor\,x^2+\sxtd\big)
\end{align*}
where $\bcor$ and $\sxtd$ are as defined in \eqref{eq:part-b-bound-params-a}  and
\begin{align*}
a = \frac{1}{2}\log_2\big((2\pi\mathrm{e})^2\srxq\sryq\big). 
\end{align*}
Due to the monotonicity of the integral and symmetry properties, we have 
\begin{align*}
\E_{X}\big[h(&\Rsym_1X+N_1|\Rsym_2)\big]\\
 &= \E_{X}\big[h(\Rsym_1X+N_1,\Rsym_2)\big] - h(\Rsym_2)\\
&\leq 2\int_{0}^{\infty}\bigg(a+\frac{1}{2}\log_2\big(\bcor\,x^2+\sxtd\big)\bigg)f_X(x)\,\mathrm{d}x - h(\Rsym_2).
\end{align*}
To evaluate the integral, we use the substitution $u=x^2$, the correspondence \cite[2.6.23.4]{Prudnikov1986}\footnote{Please note that in \cite[2.6.23.4]{Prudnikov1986} the sign before ${}_2F{}_2(\cdot;\cdot;\cdot)$ is incorrect.}, and the identities $\mathrm{erfi}(y)=-\imath\,\mathrm{erf}(\imath y)$ and $\frac{\sqrt{\pi}}{2\,z}\mathrm{erf}(z)={}_1F{}_1\big(\frac{1}{2};\frac{3}{2};-z^2\big)$ \cite[13.6.7]{Olver2010}. 
Using the expression of the differential entropy $h(\Rsym_2)$ given in \eqref{eq:part-b-diff-h-r2} yields the bound for  $\E_{X}\big[h(\Rsym_1X+N_1|\Rsym_2\big)]$ given in Proposition~\ref{prop:part-b-bound}. 
\end{IEEEproof}

\section{Proof of Proposition~\ref{Prop:Part-C}}\label{app:proof-prop:Part-C}
\begin{IEEEproof}
Fix $\Rsym_1=\rsym_1$ for some $\rsym_1 \geq 0$. Then, $\rsym_1X+N_1$ is a Gaussian random variable with zero mean and variance $\rsym_1^2P+\sigma_{N_1}^2$ since $X$ and $N_1$ are independent Gaussian random variables with positive variances $P$ and  $\sigma_{N_1}^2$. For the  differential entropy $h\big(\rsym_1X+N_1)$, we obtain
\begin{align}\label{eq:part-c-ratio-represenation}
h\big(\rsym_1X+N_1)=\frac{1}{2} \log_2\big(2\pi\mathrm{e}\sigma_{N_1}^2P\big)+ \frac{1}{2} \log_2\bigg(\frac{\rsym_1^2}{\sigma_{N_1}^2}+\frac{1}{P}\bigg). 
\end{align}
Using \eqref{eq:part-c-ratio-represenation}, we can write  
\begin{align*}
\E_{\Rsym_1}&\big[h(\Rsym_1X+N_1)\big]\\
 &= h(N_1)+\frac{1}{2}\log_2\big(P\big)+\frac{1}{2}\,\E_{T_1}\big[\log_2\big(T_1+\nicefrac{1}{P}\big)\big] 
\end{align*}
where $T_1=\Rsym_1^2/\sigma_{N_1}^2$. With the marginal pdf \eqref{eq:pdf-r1} of the random variable $\Rsym_1$ and basic density transformation, we obtain the pdf
\begin{align*}
f_{T_1}(t_1)=\exp(-t_1),\quad t_1 \geq 0 
\end{align*}
of the random variable $T_1$ such that we have
\begin{align*}
\E_{T_1}\big[&\log_2\big(T_1+\nicefrac{1}{P}\big)\big]\\
 &= \int_{0}^{\infty}\log_2\big(t_1+\nicefrac{1}{P}\big)\exp(-t_1)\,\mathrm{d}t_1\\
&=-\log_2(P)-\frac{1}{\ln(2)}\exp\bigg(\frac{1}{P}\bigg)\mathrm{Ei}\bigg(-\frac{1}{P}\bigg).
\end{align*}
The integral is solved using the substitution $u=(t_1+\nicefrac{1}{P})$ and integration by parts. Collecting the terms yields \eqref{eq:Prop:Part-C}. 
\end{IEEEproof}
%

\section{Figures for Numerical Results}\label{app:figures}
See Figs.\,\ref{fig:bounds-rho-01}-\ref{fig:bounds-rho-09} below. 

\begin{figure*}[tb]
\centering
\includegraphics[width=\textwidth]{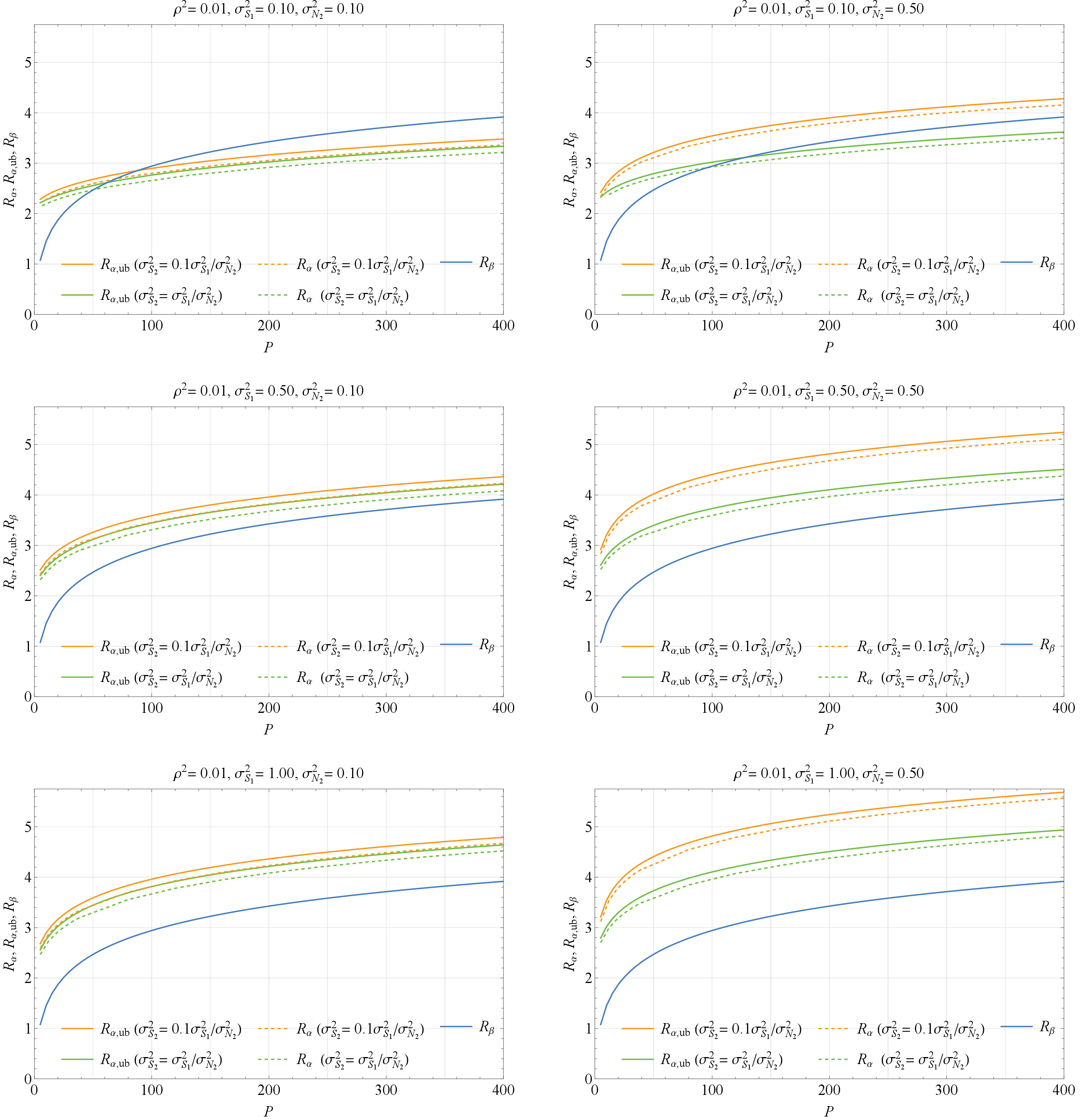}
\caption{$\Ralpha,\RalphaUB,\text{ and }\Rbeta$ for power correlation coefficient $\fcor=0.01$.}
\label{fig:bounds-rho-01}
\end{figure*}

\begin{figure*}[tb]
\centering
\includegraphics[width=\textwidth]{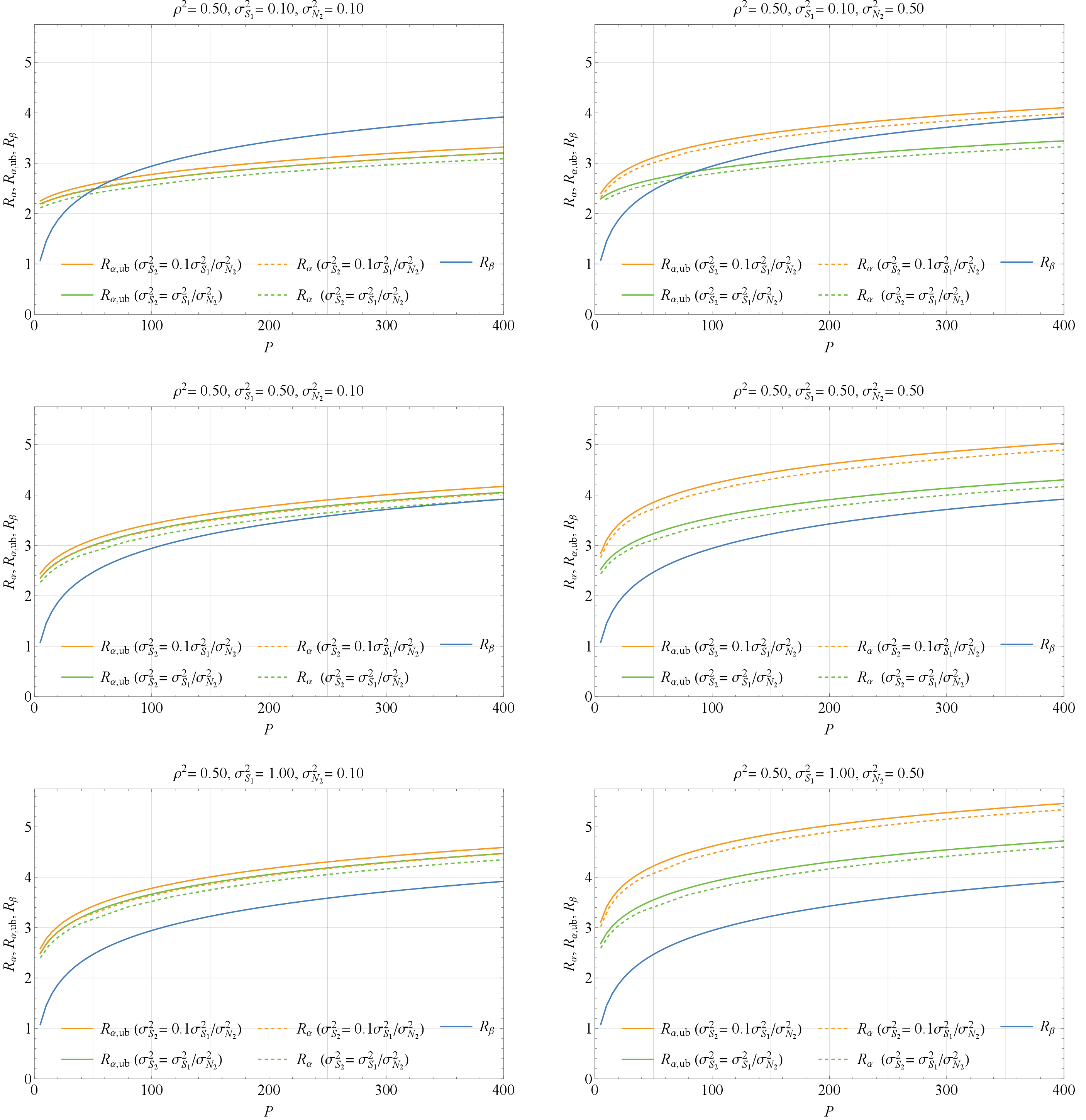}
\caption{$\Ralpha,\RalphaUB,\text{ and }\Rbeta$ for power correlation coefficient $\fcor=0.50$.}
\label{fig:bounds-rho-05}
\end{figure*}

\begin{figure*}[tb]
\centering
\includegraphics[width=\textwidth]{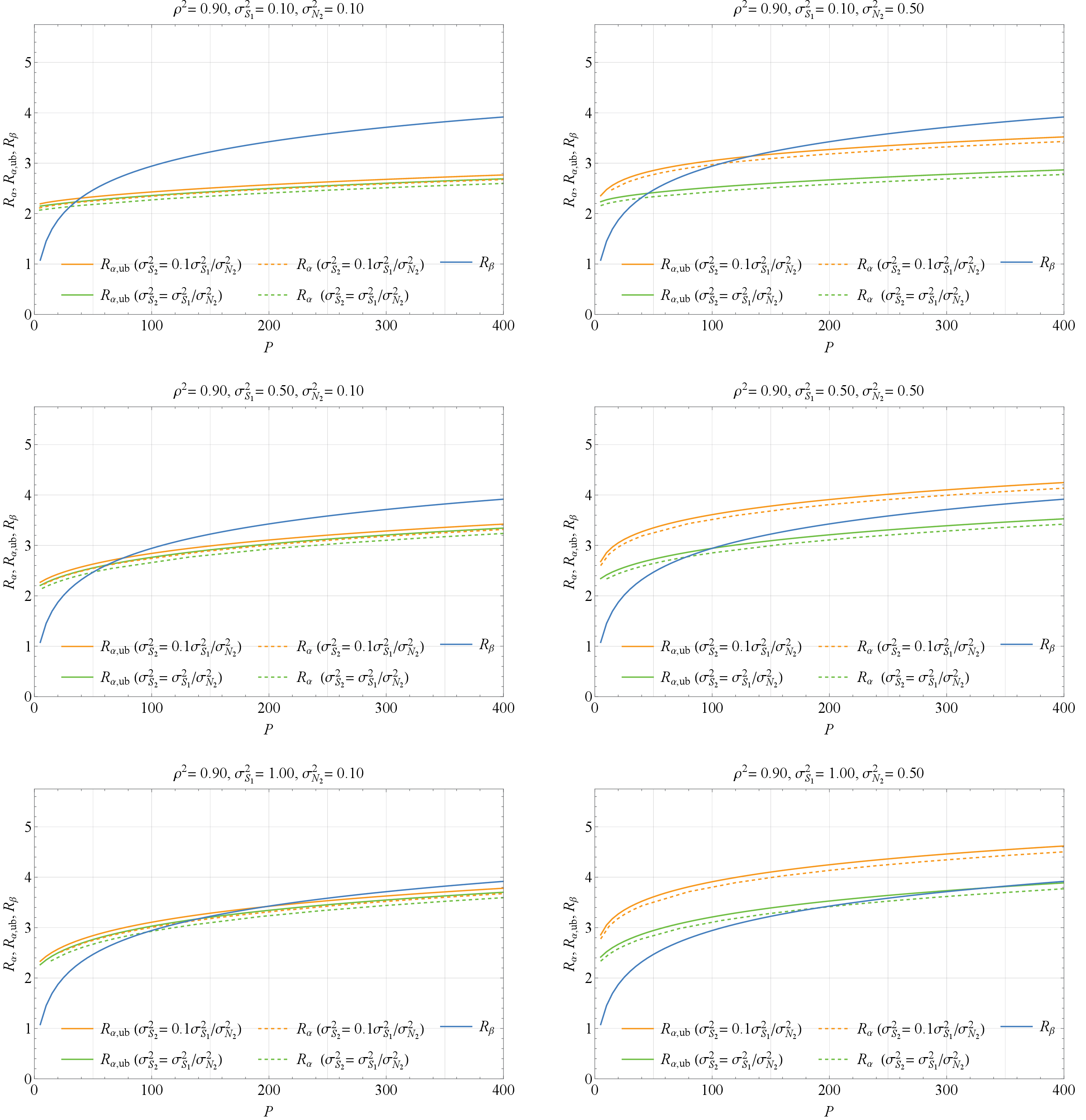}
\caption{$\Ralpha,\RalphaUB,\text{ and }\Rbeta$ for power correlation coefficient  $\fcor=0.90$.}
\label{fig:bounds-rho-09}
\end{figure*}

\end{document}